# SERS discrimination of single amino acid residue in single peptide by plasmonic nanocavities

*Jian-An Huang [a]\*, Mansoureh Z. Mousavi [a], Giorgia Giovannini [a], Yingqi Zhao [a], Aliaksandr Hubarevich [a], Denis Garoli [a,b]\*, Francesco De Angelis [a]*

[a]Istituto Italiano di Tecnologia, Via Morego 30, 16163 Genova, Italy
[b]AB ANALITICA s.r.l., Via Svizzera 16, 35127 Padova, Italy
*Email: Jian-an.huang@iit.it, Denis.garoli@iit.it

## Abstract

Surface-enhanced Raman spectroscopy (SERS) is a sensitive label-free optical method that can provide fingerprint Raman spectra of biomolecules such as DNA, amino acids and proteins. While SERS of single DNA molecule has been recently demonstrated, Raman analysis of single protein sequence was not possible because the SERS spectra of proteins are usually dominated by signals of aromatic amino acid residues. Here, we used electro-plasmonic approach to trap single gold nanoparticle in a nanohole for generating a plasmonic nanocavity between the trapped nanoparticle and the nanopore wall. The giant field generated in the nanocavity was so sensitive and localized that it enables SERS discrimination of 10 distinct amino acids at single-molecule level. The obtained spectra are used to analyze the spectra of 2 biomarkers (Vasopressin and Oxytocin) made of a short sequence of 9 amino-acids. Significantly, we demonstrated identification of single non-aromatic amino acid residues in a single short peptide chain as well as discrimination between two peptides with sequences distinguishable in 2 specific amino-acids. Our result demonstrate the high sensitivity of our method to identify single amino acid residue in a protein chain and a potential for further applications in proteomics and single-protein sequencing.

# Introduction

The advance of analytical techniques with extremely low limits of detection has led to dramatic progresses in the field of single molecule sensing and in particular in nucleic acids sequencing. While revolutionary advances in DNA sequencing technology have been achieved during the last decade,[1] another technological revolution would be a fast and reliable analysis and sequencing of proteins,[2, 3] which are the primary actors in virtually all life processes and are coded by DNA sequences known as genes. Proteins play vital functional and structural role in living systems and in-depth investigation of proteins is important in the development of proteomics. Yet, in protein biomarker discovery, it is difficult to sequence low abundance protein molecules in a complex matrix, which also contain post-translational protein modifications.[4] Single-molecule analysis of proteins would be of enormous value by offering the potential to detect extremely small quantities of proteins that may have been altered by alternative splicing or post-translational modification.

Conventional approaches for detection and characterization of proteins include mass spectroscopy (MS), X-ray crystallography and nuclear magnetic resonance (NMR). In particular, now the analysis of the protein sequence requires complicated and expensive procedure based on the MS. To enable rapid and low-cost protein analysis, innovative analytical methods have been proposed during the recent years.[2] Among the others, one of the most promising is represented by Surface-enhanced Raman spectroscopy (SERS) that enables label-free detections of analyte molecules by their fingerprint Raman spectra of biomolecules, such as DNA and proteins, with single-molecule sensitivity.[5] SERS relies on enhanced and localized electromagnetic field (so-called hot spot) on plasmonic metal nanostructures, such as gold nanoparticles, that can concentrate light on the nanoparticle surface upon illumination by laser of resonant wavelength. If the analyte molecules are adsorbed on the hot spot, they will be excited by the enhanced field and emit the SERS signals. The electromagnetic field at a nanoscale gap between two neighbouring gold nanoparticles can be even more enhanced and localized due to coupling effect so single-molecule SERS is possible.[6]

Although various SERS substrates have been developed, label-free SERS of proteins faces three major challenges: native state, sensitivity and signal reproducibility.[7, 8] The native states of proteins are related to structural functions and should be reduced to the primary state during the detection process. For example, the proteins should be measured in aqueous physiological environment rather than dry states. Moreover, proteins diffusion in solution would exhibit a few different conformations and orientations in the hot spots. More importantly, protein molecules were so large that only part of the proteins, namely amino acid residues, were adsorbed in the hot spots. These two factors gave rise to SERS signal fluctuations.[9] The challenge of size difference between the protein molecules and the hot spots could be circumvented by spectral analysis of the SERS spectra of the amino acid residues.[10] Accordingly, many reports focused on SERS detections of amino acids[11, 12], dipeptides[10, 13], polypeptides and their analogs with sequence difference of 2 amino acids[14-16]. Significantly, Clement *et al*. introduced an analysis method to retrieve conformations of bovine serum albumin adsorbed on gold nanoparticles in a microfluidic channel by analyzing the SERS spectra of single or a few amino acid residues of the bovine serum albumin molecules.[9] To allow reliable analysis, advanced plasmonic colloids systems were developed in recent years to produce stable SERS spectra of proteins and amino acids.[7, 17, 18]

Nevertheless, the SERS spectra of proteins generally exhibited broad peaks due to adsorption of many amino acid molecules in the hot spots.[10] Meanwhile, they were dominated by the spectra of aromatic amino acid residues,[19] because Raman cross-sections of the aromatic amino acids were 10 - 1000 times larger than the Raman cross-sections of the non-aromatic amino acids.[20] This actually leads to a long-standing challenge to SERS detections of proteins. Among 21 types of amino acids, only 4 of them have aromatic rings. Although non-aromatic amino acids can be detected by SERS, they were invisible in the polypeptide spectra. As a result, plenty structural information of the proteins were missing unless single-molecule spectra of non-aromatic amino acid residues in a protein were obtained. More importantly, these single-molecule spectra can lay a foundation for development of single-molecule protein sequencing, which is challenging but can find broad applications to both basic research and clinical dignostics.[2]

In this work, we used an electroplasmonic trapping system (Figure 1a) to collect and demonstrate distinguishable single-molecule SERS spectra of 10 different types of amino acids, with or without aromatic rings. The chosen amino-acids are the fundamental building blocks of 2 important biomarkers: Oxytocin (OXT) and Vasepressin (VAS), as shown in Figure 1b,c.[21] Encouraged by small peak widths of the single-molecule spectra, we further measured OXT and VAS at single-molecule level and identified successfully non-aromatic amino acid residues from aromatic ones in their structures to discern their sequence difference of 2 amino acids residues by multiplexing SERS spectra.

To this aim, sub-monolayer of the molecules were firstly adsorbed on gold nanourchins (AuNUs) before the AuNUs were driven to a gold nanohole by electrical bias. Due to a balance between electrophoresis force and electroosmosic force exerted on the AuNU, the AuNU could stay in the nanohole for tens of seconds and was pulled to the nanohole wall by an optical force due to plasmonic resonance of the nanohole with illuminating laser.[22] The nanogap formed between the AuNU tip and the nanohole wall exhibited giantly-enhanced electromagnetic field (Figure 1a Inset) with a nanoscale volume (so-called plasmonic nanocavity)[23] that continuously excited the adsorbed molecules on the AuNU tip to produce reproducible single-molecule SERS spectra, as already demonstrated by our previous work.[22] By addressing the 3 challenges in protein SERS mentioned above, we believe that this approach can represent a significant improvement towards the label-free protein sensing and single-molecule protein sequencing.

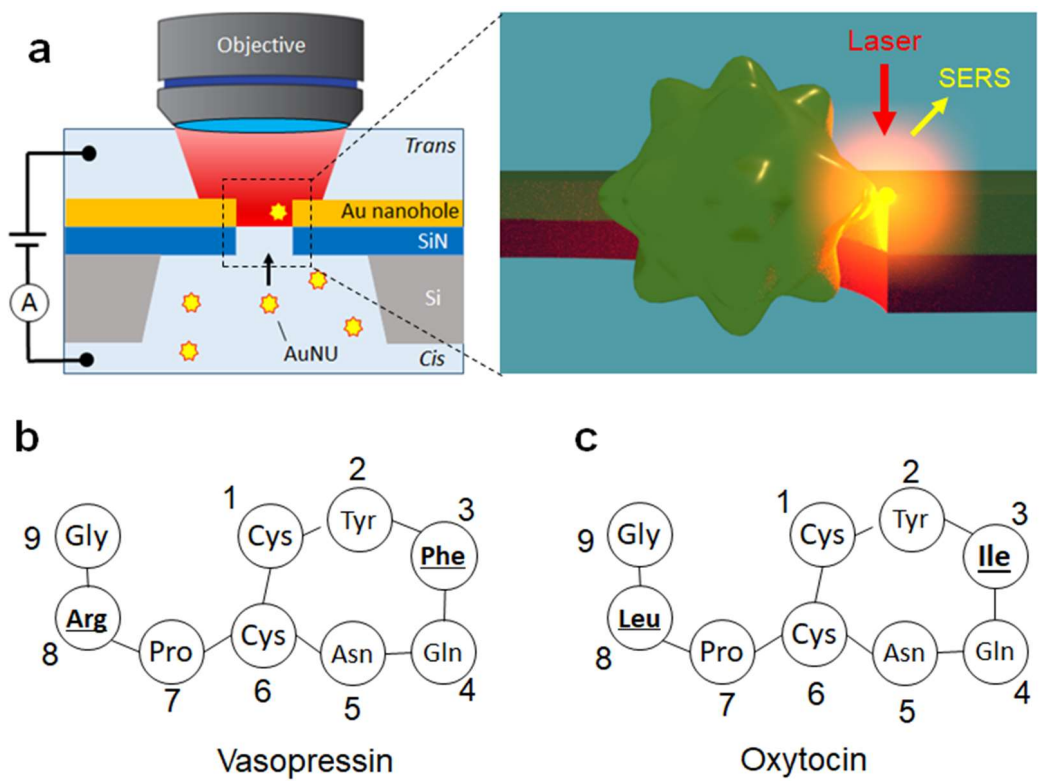

*Figure 1. (a) Schematic of the flow-through setup that allows single AuNUs to flow through and be trapped in a gold nanohole with plasmonic resonance upon the laser excitation at 785 nm. Inset: the trapping leads to a plasmonic hot spot between the AuNU tip and the nanohole sidewall that allows single-molecule SERS. The sequences of the 2 polypeptides measured by the electro-plasmonic system: (b) Vasopressin and (c) Oxytocin, in which the different amino acid residues are underlined.*

## Results and Discussions

### Single-molecule spectra of amino acids

An AuNU has many sharp tips on which incident electromagnetic field could be concentrated as hot spots. When multilayer of amino-acid molecules, like Cystine (Cys), were adsorbed on the AuNUs (Cys-AuNU), the SERS spectra from Cys-AuNUs diffusing in solution were multi-molecule spectra from many hot spots (black curve in Figure 2b). These spectra were not stable in time because they are generated by multiple hot-spots from AuNU that can also form clusters during their diffusion (Supplementary Figure S1)

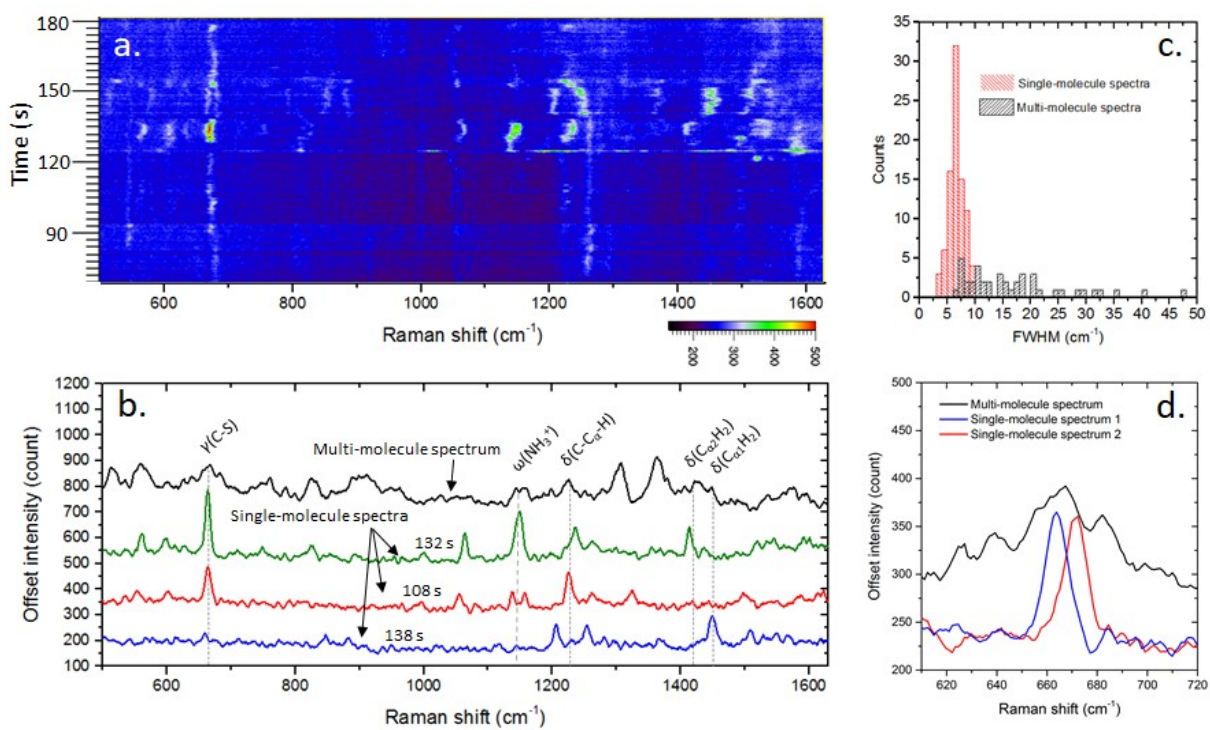

*Figure 2. (a) SERS time series of trapped submonolayer Cys-AuNU. The color bar represent peak intensity. (b, d) Comparisons between single-molecule spectra (color lines) of trapped submonolayer Cys-AuNU extracted from (a) and multi-molecule spectrum (black line) from multilayer Cys-AuNU diffusion in solution (Supplementary Figure S1). The vertical grey dashed lines indicate assignment of Cys SERS modes according to Ref.[13] (c) Histograms of the γ(C-S) peak widths (FWHM) of the single-molecule spectra and multi-molecule spectra that were fitted to give an average width as 6.6 and 17.4 $cm^{-1}$, respectively.*

On the contrary, a Cys-AuNU trapped in the plasmonic nanohole for tens of seconds actually generated single hot spot by the nanogap between the AuNU tip and the nanohole wall.[22] When we lowered the number of the adsorbed Cys molecules on the AuNUs to the level of sub-monolayer (see details in Supporting Information), the Cys sub-monolayer allowed single Cys molecule in the hot spot to produce stable single-molecule spectra with low-noise baselines and narrow peaks (Figure 2a).[22] In contrast to an average peak width of 17.4 $cm^{-1}$ for the multi-molecule spectra (Figure 2b – black curve), the single-molecule peaks had an average peaks width as small as 6.6 $cm^{-1}$ (Figure 2c). In fact, the broad multi-molecule peak, such as the γ(C-S) mode in Figure 2d, were a sum of the shifting single-molecule peaks,[24] each of which represented specific conformations and orientation of the adsorbed Cys molecules.[25] We observed also peak blinking, such as the δ($C_{α2}H_2$) mode at 132 s and the δ($C_{α1}H_2$) mode at 138 s in Figure 2b, which could be ascribed to translational and/or rotational movement of the Cys molecule.[26] Therefore, within the stable single hot spot, molecule dynamics can be monitored by the single-molecule spectra. For example, the Cys stranded on the AuNU tip surface with a trans position of the -$CH_2$-S- bond after 156 s because only the $\gamma$(C-S) mode remained active (Figure 2a).[13]

To test the sensitivity of the platform, we collected single-molecule spectra of additional 9 different amino acids, including aromatic Phenylalanine (Phe), Tyrosine (Tyr) as well as non-aromatic Glycine (Gly), Leucine (Leu), Isoleucine (Ile), Arginine (Arg), Proline (Pro), Glutamine (Gln), Asparagine (Asn). Together with the Cys, they are the building blocks of the OXT and VAS. Their reproducible single-molecule spectra also exhibited similar narrow peaks and low-noise baseline as those of the Cys (Supplementary Figure S2 – S7). Successful identification of the amino acids were performed by assigning at least 2 peaks (see Supplementary Table S3 for peak assignments).[11] The small peak widths shows the potential of using the multiplexing single-molecule spectra to discern single amino acid residues in the OXT and VAS.

**Discrimination of single amino acids in a single peptide**

Both OXT and VAS are peptide of nine amino acids (a nonapeptide). Their sequence are: H-Cys-Tyr-**Ile**-Gln-Asn-Cys-Pro-**Leu**–Gly-$NH_2$ and H-Cys-Tyr-**Phe**-Gln-Asn-Cys-Pro-**Arg**-Gly-$NH_2$, respectively. As underlined, the sequences of VAS and OXT differ by 2 amino acid residues at positions 3 (Ile → Phe) and 8 (Leu→Arg). However, since the VAS has the Tyr and Phe residues and the OXT has the Tyr residue, the multi-molecule SERS spectra of the peptides were usually overwhelmed by the spectra of the corresponding aromatic amino acids.[14, 19] Discrimination of their sequence difference were not possible until single-molecule spectra of non-aromatic amino acid residues were identified.

By trapping AuNUs coated respectively with VAS and OXT submonolayers (VAS-AuNU and OXT-AuNU), we observed spectra down to single-residues per time. For example, Phe were observed in measuring a VAS-AuNU (Figure 3a). Interestingly, blinking of the Phe ring mode, CC(ring) at 1006 cm$^{-1}$, suggested that the VAS side chain was changing. Furthermore, discrimination of residue difference at the 3$^{rd}$ position near the Tyr residue in the 2 peptides was more challenging, because the aromatic Tyr may produce strong SERS signal to hide its neighbors. Yet impressively, nonaromatic Ile and aromatic Tyr were identified in measuring a OXT-AuNU at single-molecule level for the first time (Figure 3b). The 2-amino-acid-residues spectra were still dominant, even though we decreased the number of the respective peptides adsorbed on the AuNUs further by half (sOXT-AuNU and sVAS-AuNU in Supplenmentary S1). While our system has demonstrated a size of the hot spot down to single nucleotide in our previous work[22], the large probability of this 2-spectra phenomena could be due to folding of the neighboring residues of the peptides. The hot spot size can be estimated as 1.7 nm by maximum accessible surface areas of Phe (2.28 nm$^2$ from Supplementary Table S1).

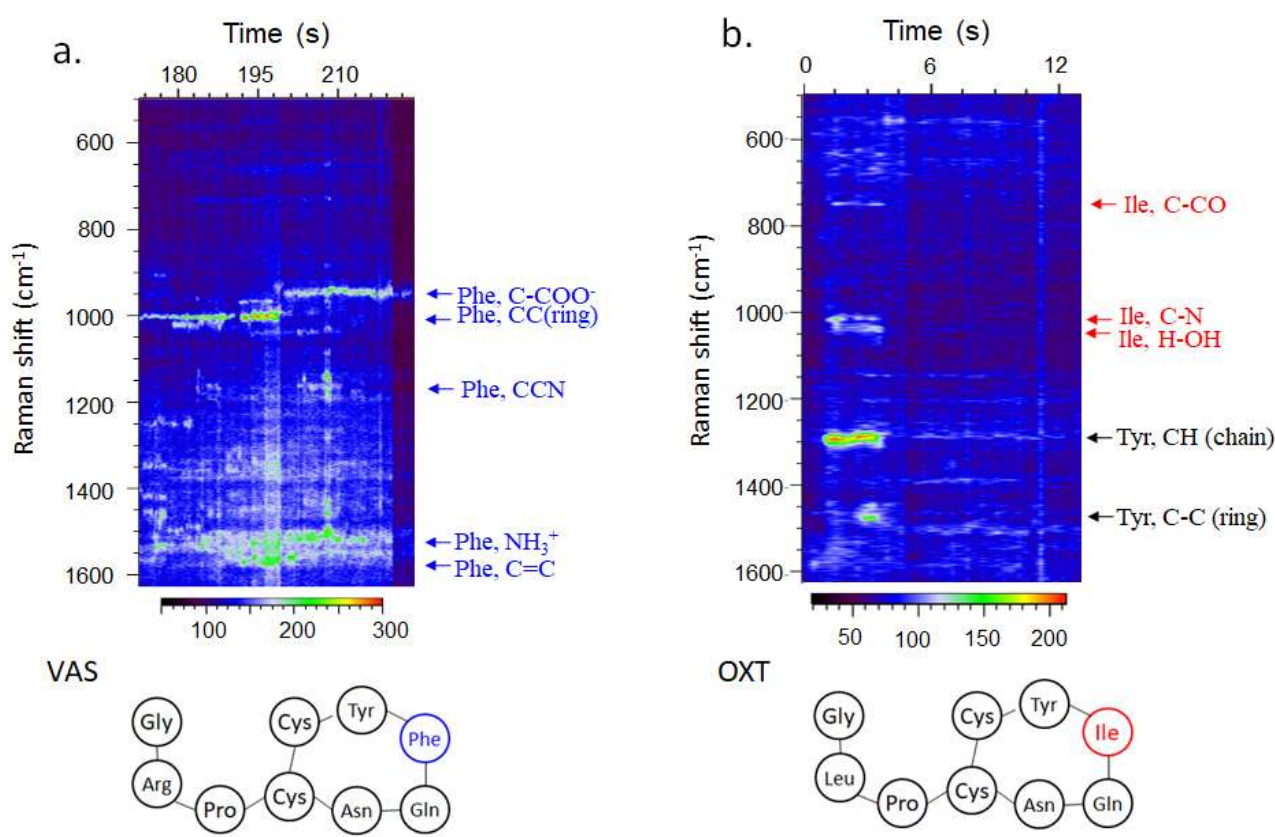

*Figure 3. SERS time series of (a) Phe of single VAS-AuNU. (b) Ile and Tyr of single sOXT-AuNU. Arrows indicate assignment of the SERS modes of corresponding amino acids. Color bars represent signal-to-baseline intensities.*

The small peak width of the peptide spectra were beneficial to multiplexing discrimination. For example, the Arg, β(NH$_2$) mode and the Pro, ω(NH$_3^+$) mode that were separated by around 20 cm$^{-1}$ (Figure 4a). Moreover, the detected Pro modes in the VAS-AuNU spectra were different from that in the OXT-AuNU spectra, such as the Pro, νCC (ring) mode in Figure 4b. They are also different from the single-molecule spectra of single Pro molecule adsorbed on the AuNU tip (Supplementary Figure S7). The differences suggested different Pro orientations due to spatial occupancy of the neighboring amino acid residues, which inversely confirmed that the 2-molecule spectra were actually from single amino acid molecules.

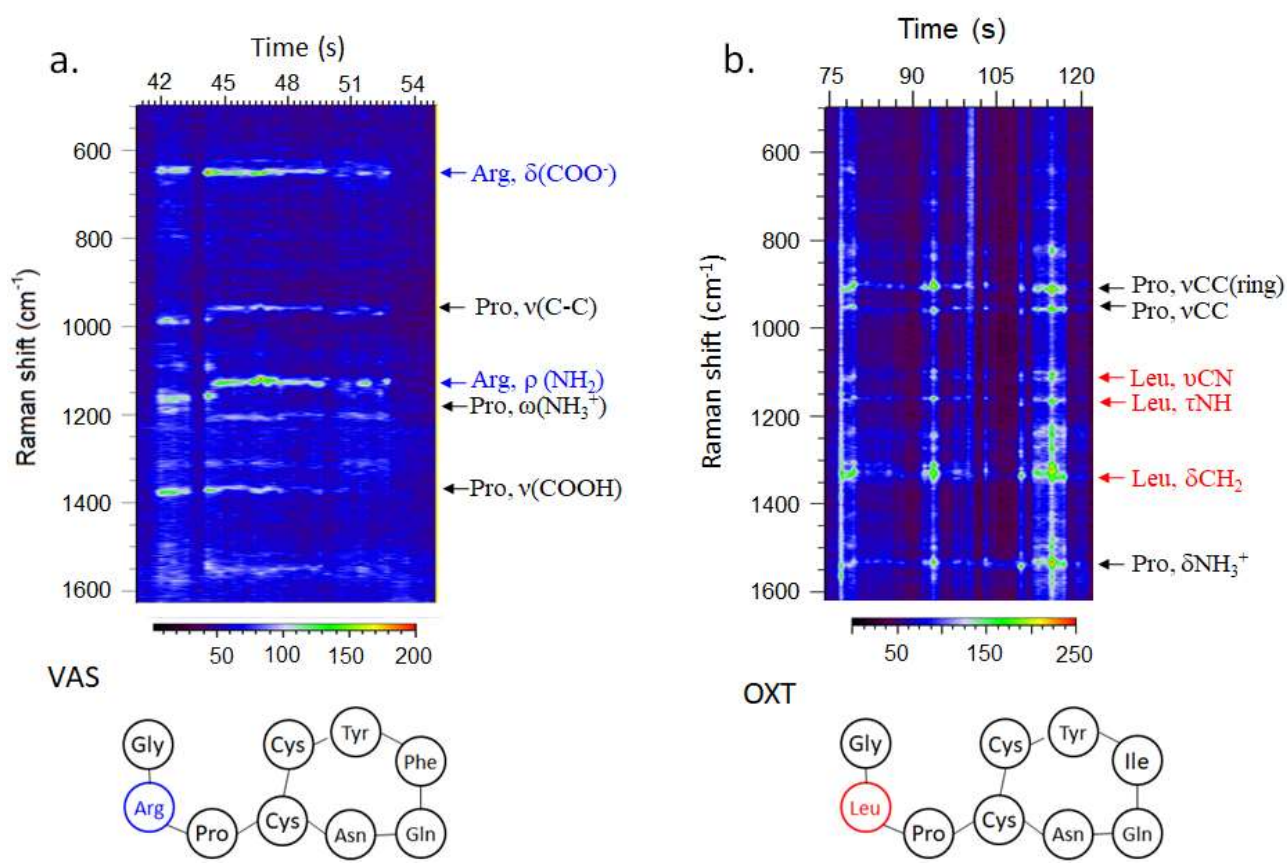

*Figure 4. SERS time series of (a) Pro and Arg of single VAS-AuNU. (b) Pro and Leu of single OXT-AuNU. Arrows indicate the SERS modes of corresponding amino acids. Color bars represent signal-to-baseline intensities.*

The discrimination of different types of amino acid residues in single peptide can contribute to single-protein sequencing. Our case of discerning 2 types of amino acids in the OXT or VAS consisting of 9 amino acids can read $2^9 = 512$ unique protein sequences.[27] We also found small amount of 3-molecule spectra in measuring the sOXT-AuNU that discerned the neighboring Asn, Gln and Ile residues by the multiplexing SERS peaks (Figure 5). Such 3-molecule spectra from a peptides consisting of 9 residues can enlarge the number of protein sequences up to $3^9 = 19,683$. To increase the probability of detecting more types of amino acid residues, we just need to trap a gold nanosphere to obtain a large hot spot. Therefore, our method show a potential to sequencing single proteins.

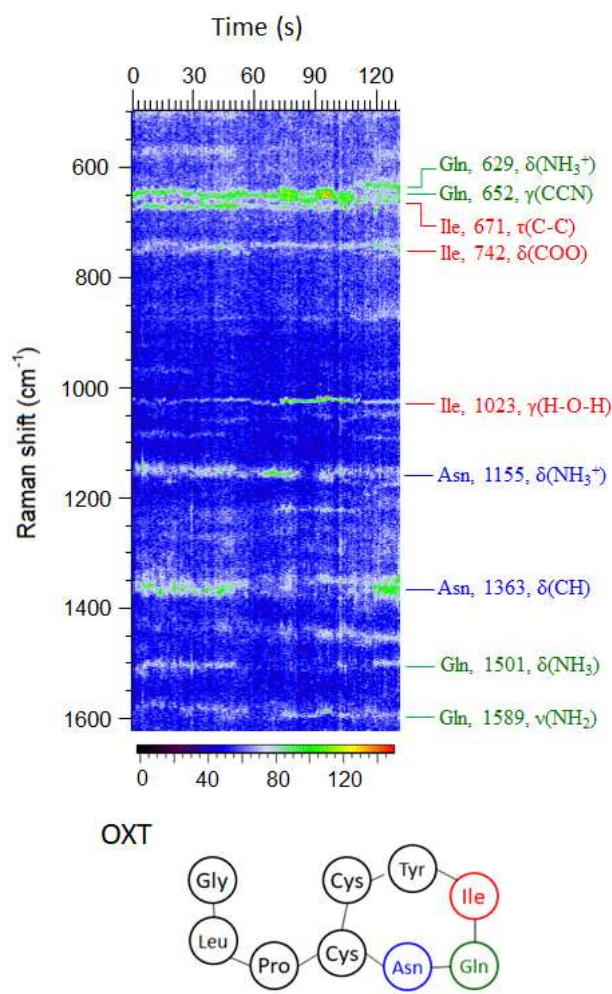

*Figure 5. SERS time series of Gln, Ile and Gln of sOXT-AuNU. Color lines indicate assignment of the SERS modes of corresponding amino acids; the numbers are the Raman shift frequency with unit of cm$^{-1}$. Color bars represent signal-to-baseline intensities.*

## Conclusion

By forming stable nanogap with localized and strong hot spots after molecule adsorption, our approach exhibited both high signal reproducibility and single-molecule sensitivity. Among the 10 amino acids, those without aromatic rings were detected at single-molecule level for the first time. Furthermore, the width of single-molecule peaks were so narrow that the sequence difference by 2 amino acids between single OXT and single VAS were also identified for the first time. The time-resolved SERS spectra of single molecules were reproducible for tens of second that protein dynamics could be monitored in real-time by our platform. By meeting the 3 challenges in protein SERS mentioned above, we believe that this approach can represent a significant improvement towards the label-free protein sensing and single-molecule protein sequencing.

# Methods

**Materials.** Non-functionalized Gold Nano-Urchins (AuNUs) with average particle sizes of 50 nm and concentrations of $3.5 \times 10^{10}$ particles/mL were obtained from Sigma (795380-25ML). The amino acids, Oxytocin (O3251) and Vasopressin (V0377) were purchased from Sigma. Phosphate Buffered Saline (806552 Sigma) was used for preparation of the samples and Raman measurements.

**Fabrication of the nanohole devices and PDMS encapsulation.** After sputtering a 2 nm thick titanium and 100 nm thick gold layer on the front side of the $Si_3N_4$ membrane, as well as a 2 nm-thick titanium and 20 nm thick gold layer on its back side, focused ion beam milling (FIB, FEI Helios NanoLab 650 DualBeam) at a voltage of 30 keV and a current from 0.23 to 2.5 nA was used to drill hole arrays in the back of the Ti/Au-coated $Si_3N_4$ sample. The sample was annealed on a hot plate at 200°C in air for 1 hour and allowed to cool naturally. The as-made nanoholes were embedded in a microfluidic chamber made from polydimethylsiloxane (PDMS, Dow Corning SYLGARD 184 silicone elastomer) cured at 65°C for approximately 40 min.

**Raman measurements.** Raman measurements were obtained by a Renishaw inVia Raman spectrometer with a Nikon 60 × water immersion objective with a 1.0 NA delivering a 785-nm laser beam and an exposure time of 0.1 s. The laser beam was focused to a spot diameter of 1.5 μm with a power varying from 2 to 12 mW.

**Single-molecule data processing.** Spectra were processed using custom python scripts according to Chen et al..[28] Data were selected in a 500-1000 cm$^{-1}$ window and smoothed using a Savitzky-Golay filter. A baseline was fitted to each spectrum using 5th-order polynomial functions and subsequently removed from the spectra. Peaks were detected on the resulting spectra using pythons' signal_find_peaks_cwt function. Final peaks were selected if the peak height exceeds 4 standard deviations of the spectrum.

# Additional Information

**Supplementary information is available.**

# Acknowledgments

The authors thank Dr. M. Dipalo for drawing of schematic figures. The research leading to these results has received funding from the Horizon 2020 Program, FET-Open: PROSEQO, Grant Agreement no. [687089].

# Author Contributions

D.G., M.Z.M., F.D.A. and J.A.H. conceived the project. J.A.H. designed the experiment, carried out Raman experiments, and wrote the manuscript. M.Z.M. and G.G. handled the molecules' attachment on the AuNUs and the AuNU stability. G.G. measured the Zeta potentials and absorbance of the functioned AuNUs. Y.Z. fabricated the nanohole device. J.A.H and A.H. analyzed the data. F.D.A. and D.G. supervised the work. All authors discussed the results and contributed to the final manuscript preparation.

Supporting information for

# SERS discrimination of single amino acid residue in single peptide by plasmonic nanocavities

*Jian-An Huang [a]\*, Mansoureh Z. Mousavi [a], Giorgia Giovannini [a], Yingqi Zhao [a], Aliaksandr Hubarevich [a], Denis Garoli [a,b]\*, Francesco De Angelis [a]*

[a]Istituto Italiano di Tecnologia, Via Morego 30, 16163 Genova, Italy

[b]AB ANALITICA s.r.l., Via Svizzera 16, 35127 Padova, Italy

*Email: Jian-an.huang@iit.it, Denis.garoli@iit.it

# S1. Sub-monolayer attachment of amino acids and polypeptides on AuNUs.

To form sub-monolayers on the AuNUs, we used concentrations of amino acids (AAs) or polypeptides that formed a monolayer on a gold nanosphere (AuNS) with the same diameter (50 nm) as the AuNUs, because the surface area of AuNS was estimated to be 400 times smaller than that of the AuNU.[29] The concentration of AA required to achieve a monolayers on the AuNUs surface was determined by empirical values for maximum solvent accessible surface area found in literature[30] and showed in the Table S1 below. The amount of polypeptide, Vasopressin (VAS) and Oxytocin (OXT), required to form a monolayer on an AuNU was determined by a surface area of a unfolded protein (in the unit of Å$^2$) as $At = 1.45 \times MW + 21$, where $MW$ is the molecular weight of the protein (in the unit of Dalton).[31] This formula is proposed for unfolded proteins as likely in the case for the small polypeptide here, and it considers the protein in its extended conformation that is the expected conformation on the AuNUs. For this assumption, the final $At$ determined with the formula corresponds to the sum of the surface area of individual residues presented in Table S1. The surface area of a single ϕ50 nm AuNUs is calculated as 7850 nm$^2$ by regarding it as a ϕ50 nm AuNS. Knowing the amount of AAs/polypeptides molecules required to form a monolayer on a single AuNUs, we then calculate the μM of molecules needed to form a monolayer on AuNUs with a final AuNU concentration of $1.3 \times 10^{10}$ mL$^{-1}$.

*Table S1: Surface area of AAs and polypeptides, the number of molecule per AuNU and final concentrations calculated for each molecule to have a submonolayer of molecule on the AuNU surface.*

| Amino acids | maximum accessible Surface area (Å$^2$) | maximum accessible Surface area (nm$^2$) | Number of molecule per AuNU | Final concentration (nM) |
|---|---|---|---|---|
| **Arg** | 265 | 2.65 | 2962 | 61.48 |
| **Asn** | 187 | 1.87 | 4198 | 87.12 |
| **Cys** | 148 | 1.48 | 5304 | 110.08 |
| **Gln** | 214 | 2.14 | 3668 | 76.13 |
| **Gly** | 97 | 0.97 | 8093 | 167.96 |
| **Ile** | 195 | 1.95 | 4026 | 83.55 |
| **Leu** | 191 | 1.91 | 4110 | 85.30 |
| **Phe** | 228 | 2.28 | 3443 | 71.45 |
| **Pro** | 154 | 1.54 | 5097 | 105.79 |
| **Tyr** | 255 | 2.55 | 3078 | 63.89 |
| Polypeptide | Molecular weight (Dalton) | Surface area $At$ (nm$^2$) | Number of molecule per AuNU | Final concentration (nM) |
| **VAS** | 1084 | 15.93 | 493 | 13.29 |
| **sVAS** | 1084 | 15.93 | 493 | 6.65 |
| **OXT** | 1007 | 14.81 | 530 | 14.30 |
| **sOXT** | 1007 | 14.81 | 530 | 7.15 |

In details, 300 μL of AuNU stock solution ($3.5 \times 10^{10}$ mL$^{-1}$) were dispersed in 400 μL of 5% PBS pH 5.5. Then, 100 μL of AAs/polypeptide solution in the same buffer were added to reach the desired concentration for the monolayer formation (final volume 800 μL with $1.3 \times 10^{10}$ AuNUs mL$^{-1}$). After vortexing, the samples were left at 4°C for two days allowing the spontaneous absorption of molecules on the AuNUs surface.

## S2. Measurement of dynamic light scattering and absorbance of the colloid.

DLS experiments were performed using a Malvern Zetasizer and the measurements were evaluated using Zetasizer software. Data are reported as the average of three measurements. Particle diameter, PDI and Zeta potential were used to characterise the colloids suspension before and after functionalisation and to evaluate their stability over time. Unless otherwise mentioned, particles were analysed at a diluted concentration of $1.3 \times 10^9$ particles/mL in filtered deionised water and in the buffer solution used for their synthesis at 25°C in disposable folded capillary cells (DTS1070). Using aqueous solutions as dispersant at 25°C, 0.8872 cP and 78.5 were used as parameters for density and dielectric constant respectively during the measurements. RI 0.2 and an absorption of 3.32 were used as parameters for the analysis of gold nanomaterials. Cary300 UV-Vis (Varient Aligent) was used for the UV-Vis analysis of the colloidal suspension. The absorbance spectrums were recorded using samples at diluted concentration of $6.5 \times 10^9$ particles/mL in water or in the buffer used for the synthesis. Absorbance measurements were performed using a 300-800 nm wavelength range in a disposable plastic cuvette (1 mL maximum volume, 1 cm path distance).

Table S2. Measured Zeta potentials ($\zeta_{np}$) of AA/polypeptide coated-AuNUs solutions.

|  | Zeta potentials (mV) | | |
| --- | --- | --- | --- |
|  | $\zeta_{np}$ 1 | $\zeta_{np}$ 2 | $\zeta_{np}$ 3 |
| **UnmodifiedAuNU** | -22 | -20.5 | -23 |
|  |  |  |  |
| **Arg-AuNU** | -11.4 | -11.1 | -9.37 |
| **Asn-AuNU** | -18.5 | -17.9 | -18 |
| **Cys-AuNU** | -17.8 | -18.6 | -15.8 |
| **Gln-AuNU** | -16.8 | -16.4 | -18 |
| **Gly-AuNU** | -26.1 | -27.2 | -25 |
| **Ile-AuNU** | -12.3 | -17.8 | -18.1 |
| **Leu-AuNU** | -28.5 | -26.2 | -30.1 |
| **Phe-AuNU** | -13.2 | -10.2 | -22.7 |
| **Pro-AuNU** | -16.1 | -12.7 | -15 |
| **Tyr-AuNU** | -18.8 | -17.9 | -17.1 |
|  |  |  |  |
| **VAS-AuNU** | -0.6 | -1.7 | -0.2 |
| **sVAS-AuNU** | -26.2 | -15.6 | -20.4 |
| **OXT-AuNU** | -1.9 | -2.3 | -2.4 |
| **sOXT-AuNU** | -13.7 | -18.6 | -16.8 |

## S3. Multilayer Cysteine adsorption on AuNUs and the SERS spectra

For multilayer Cysteine adsorption, the concentration of AuNUs and the salt concentration were $1.3 \times 10^{10}$ particles/mL of AuNUs and 1.25% of phosphate buffer. The multilayer Cysteine-AuNUs were prepared by pipetting 300 μL of the AuNUs suspension from a stock in an Eppendorf tube, which was then re-suspended in 400 μL of deionised water. Cysteine solution with final concentration of 125 μM was added in the AuNU solution and allowed to adsorb on the AuNUs with continuous mixing on a shaker at room temperature for 3 hours prior to the Raman measurement.

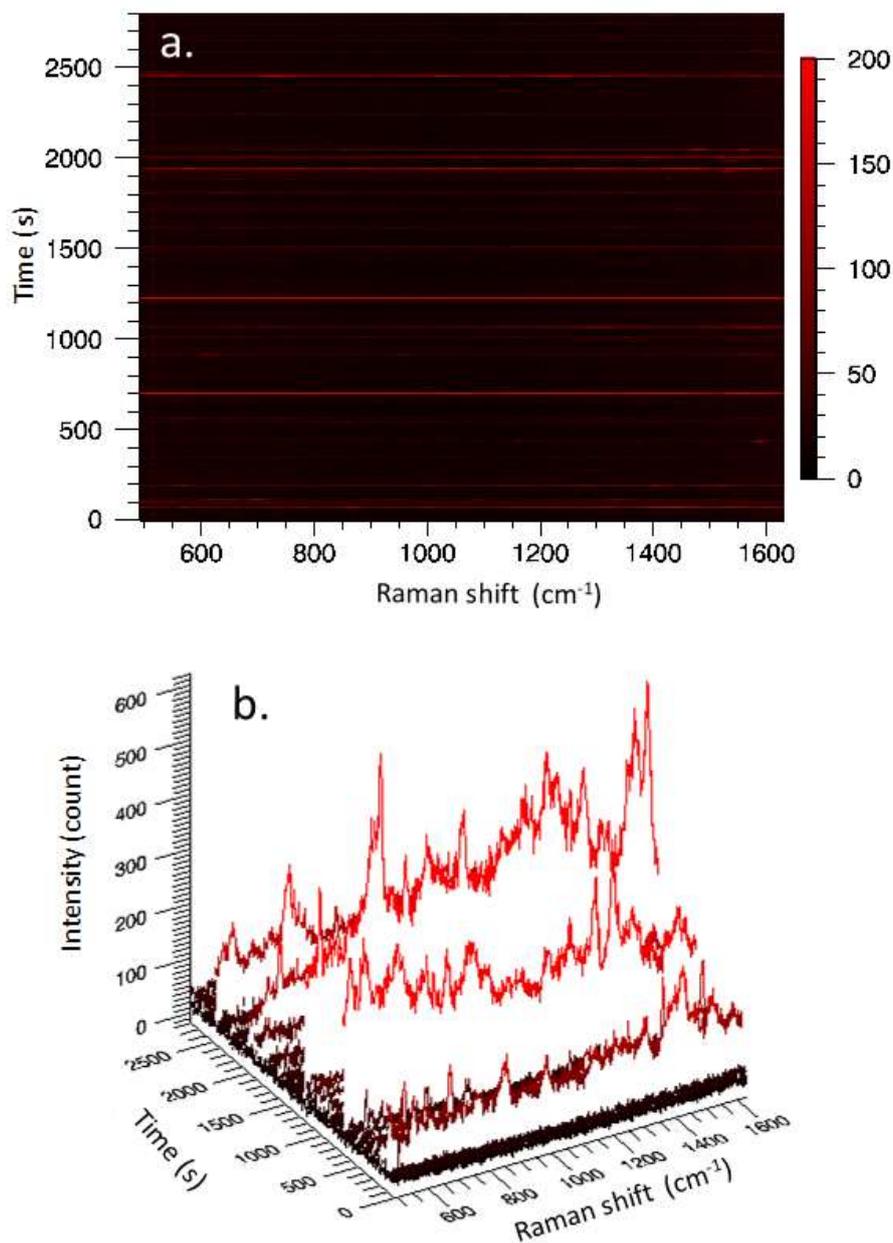

*Figure S1. SERS time series of multilayer Cys-AuNU diffusion in solution. (a) Contour view with a colorbar to represent peak intensity. (b) Spectral view in which the fluctuating spectra are visible.*

## S4. Single-molecule spectra of amino acids.

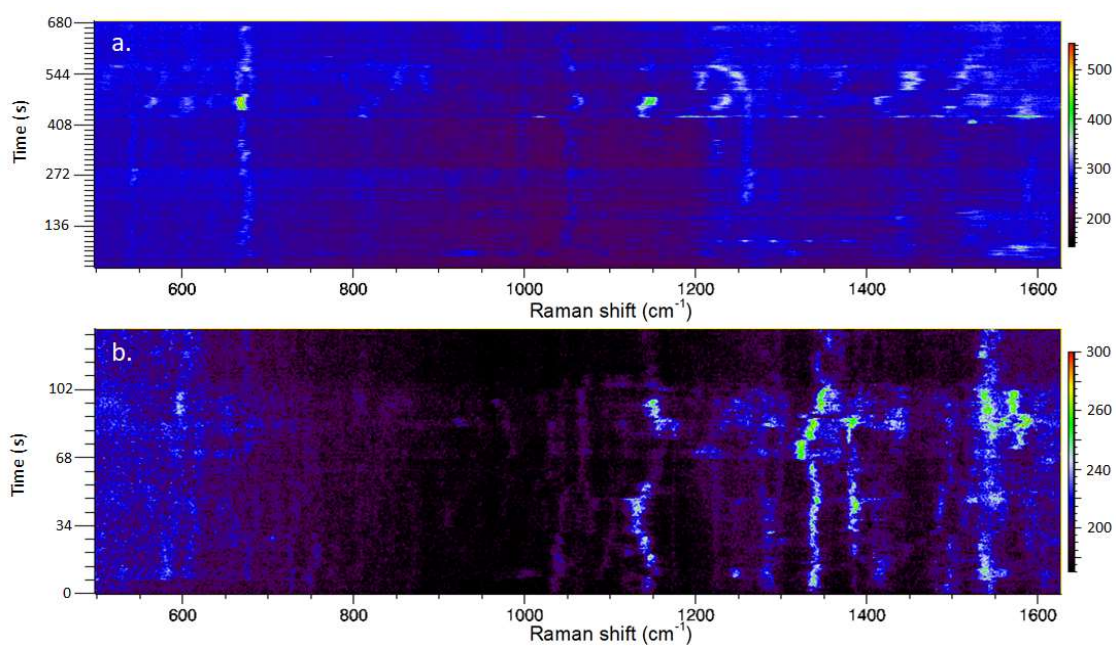

*Figure S2. SERS time series of (a) trapped submonolayer Cys-AuNU. (b) trapped submonolayer Gly-AuNU. Color bars indicate signal-to-baseline peak intensities.*

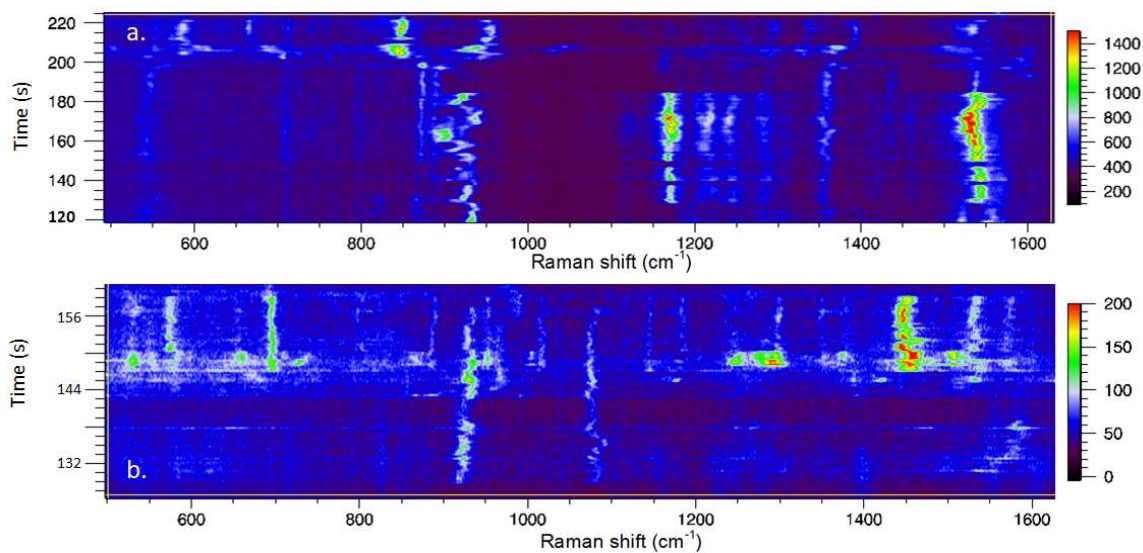

*Figure S3. SERS time series of (a) trapped submonolayer Leu-AuNU. (b) trapped submonolayer Ile-AuNU. Color bars indicate signal-to-baseline peak intensities.*

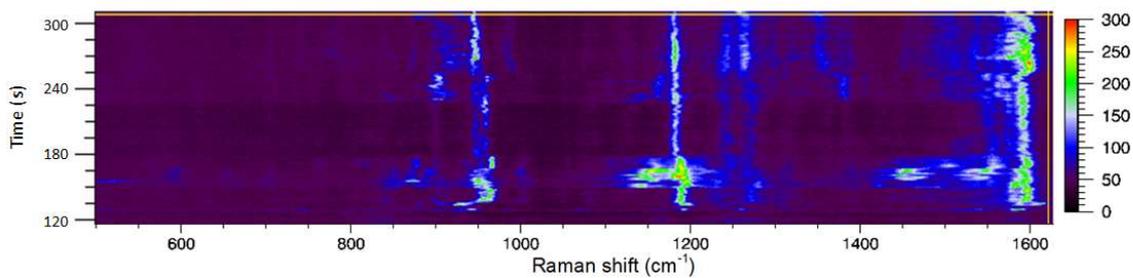

*Figure S4. SERS time series of trapped submonolayer Arg-AuNU. Color bars indicate signal-to-baseline peak intensities.*

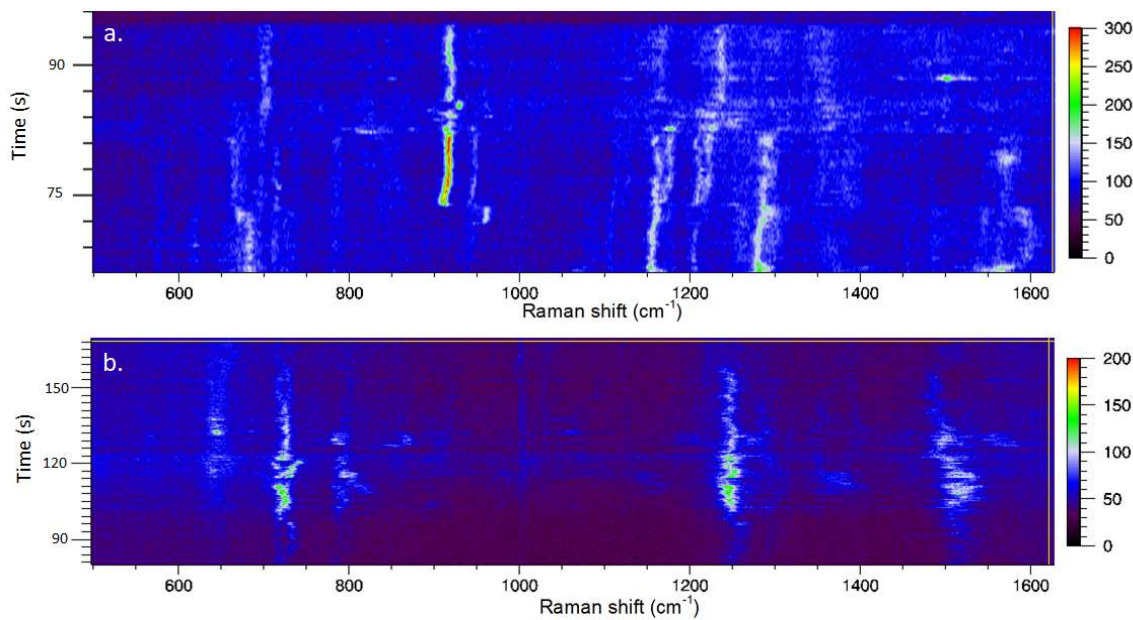

*Figure S5. SERS time series of (a) trapped submonolayer Gln-AuNU. (b) trapped submonolayer Asn-AuNU. Color bars indicate signal-to-baseline peak intensities.*

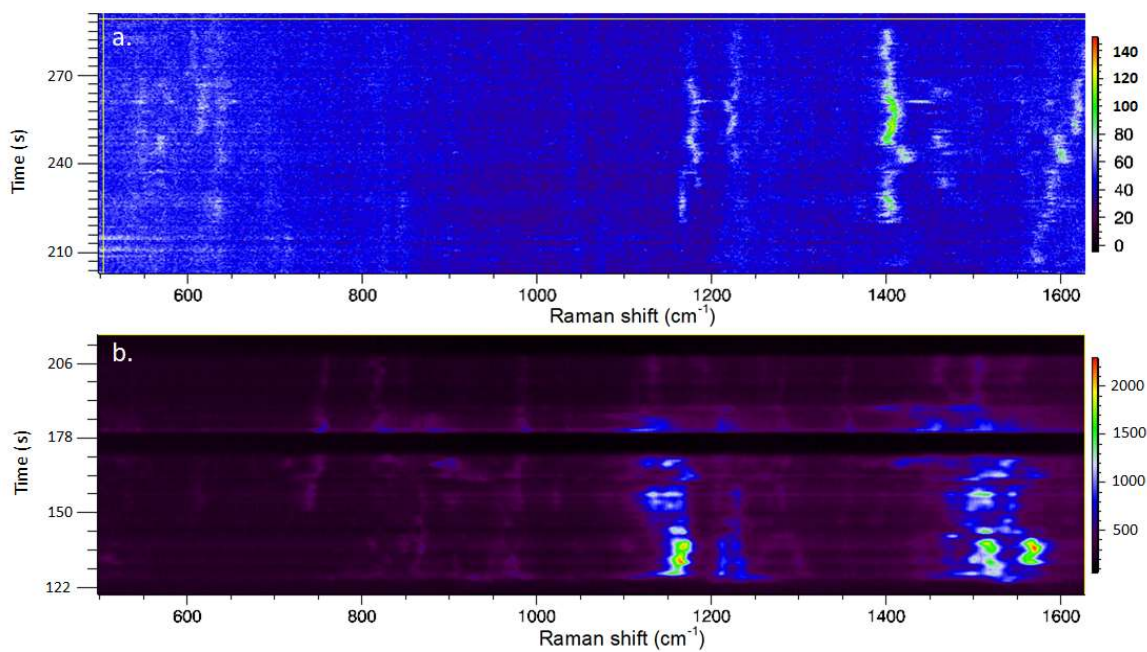

*Figure S6. SERS time series of (a) trapped submonolayer Phe-AuNU. (b) trapped submonolayer Tyr-AuNU. Color bars indicate signal-to-baseline peak intensities.*

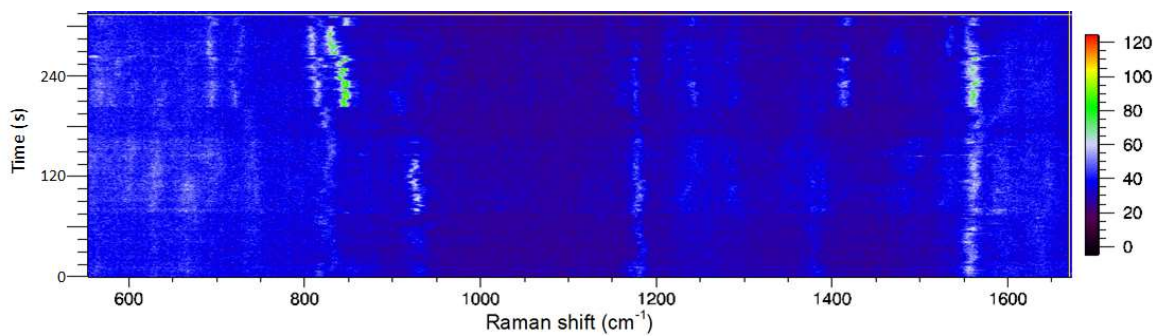

*Figure S7. SERS time series of trapped submonolayer Pro-AuNU. Color bars indicate signal-to-baseline peak intensities.*

*Table S3. Molecular structures and assignment of the SERS bands of the sub-monolayer amino aicd-AuNU in Figure S2 – S7 to specific vibrational modes.[a]*

| Amino Acid | Molecular structure | Raman modes (cm$^{-1}$) | Most probable assignment |
|---|---|---|---|
| Gly[32] | | 1139 | $\omega(C_\alpha H_2)$, $t(C_\alpha H_2)$ |
| | | 1350 | $\delta_s(N_t\text{-}C_\alpha\text{-}H)$ |
| | | 1395 | $\gamma_{as}(C_tOO^-)$ |
| | | 1540 | $\delta_s(NH_3^+)$ |
| Cys[13] | | 670 | $\gamma(C\text{-}S)$ |
| | | 1140 | $\omega(NH_3^+)$ |
| | | 1220 | $\delta(C\text{-}C_\alpha\text{-}H)$ |
| | | 1420 | $\delta(C_{\alpha 2}H_2)$ |
| | | 1451 | $\delta(C_{\alpha 1}H_2)$ |
| Leu[33] | | 850 | $\rho(CH_2)$ |
| | | 923 | $\rho(CH_3)$ |
| | | 1173 | $\nu(C\text{-}C)$ |
| | | 1537 | $\delta_s(NH_3^+)$ |
| Ile[33] | | 570 | $\rho(COOH)$ |
| | | 700 | $\nu(C\text{-}CO)$ |
| | | 930 | $\rho(CH_3)$, $\nu(C\text{-}C)$ |
| | | 1089 | $\rho(CH_3)$ |
| | | 1448 | $\delta_{as}(CH_3)$ |
| Asn[34] | | 724 | $\omega(CO_2)$ |
| | | 1249 | $\rho(NH_3)$, $\rho(CH_2)$ |
| | | 1502 | $\gamma(NH_3^+)$, $\gamma(CH)$, $\nu(CO_2)$ |
| Gln[35] | | 916 | $\nu_s(C\text{-}C)$ |
| | | 1163 | $\delta(NH_3^+)$ |
| | | 1204 | $\delta(NH_2)$ |
| | | 1283 | $\delta(NH_2)$ |
| Arg[36] | | 945 | $\rho(CH_2)$ |
| | | 1180 | $\rho(NH_3)$, $\rho(CH_2)$ |
| | | 1590 | $\delta(NH_2)$, $\delta_{as}(CH_3^+)$ |
| Pro[13] | | 844 | $\gamma(C\text{-}C)$ |

| | | 924 | γ(C-COOH) |
|---|---|---|---|
| | 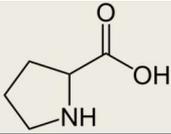 | 1180 | ω(NH$_3^+$) |
| | | 1565 | δ$_s$(NH$_3^+$) |
| Phe[13, 18] | 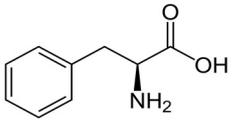 | 1160 | ω(NH$_3^+$) |
| | | 1400 | ν$_s$(COO$^-$) |
| | | 1600 | ν$_{ring}$, ν$_{as}$(COO$^-$) |
| | | 1620 | γ(NH$_2$) |
| Tyr[12, 18] | 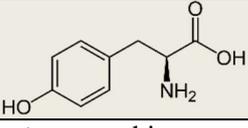 | 1163 | ρ$_{as}$ (NH$_2$) |
| | | 1520 | γ(NH), ν$_{ring}$ |
| | | 1566 | ν$_{ring}$ |

[a] Abbreviations: t: twist; ρ: rocking; ν: stretching; γ: out-of-plane bending; δ: deformation or in-plane bending; ω: wagging; τ: torsion; s: symmetric; as: antisymmetric.